\documentclass[twocolumn]{aastex62}

\usepackage{graphics,epsf}
\usepackage{amsmath}                
\usepackage{amsfonts}               
\usepackage{amssymb}                
\usepackage{epsfig}                 
\usepackage{float}
\usepackage{color}
\usepackage{tabularx}
\usepackage{comment}
\usepackage{makecell}

\def \s{~\rm{s}}
\def \km{~\rm{km}}

\def \yr{~\rm{yr}}
\def \Myr{~\rm{Myr}}
\def \Gyr{~\rm{Gyr}}

\def \kpc{~\rm{kpc}}

\usepackage{xcolor}
\definecolor{redak}{rgb}{0.9,0.15,0.05}

\begin{document}

\title{Towards a paradigm change in the main heavy r-process nucleosynthesis sites}

\author{Aldana Grichener}
\affiliation{Department of Physics, Technion, Haifa, 3200003, Israel; aldanag@campus.technion.ac.il; soker@physics.technion.ac.il}

\author[0000-0003-0375-8987]{Noam Soker}
\affiliation{Guangdong Technion Israel Institute of Technology, Shantou 515069, Guangdong Province, China}


\begin{abstract}
We study the basic properties of three possible r-process scenarios (sites) and compare them to recent observations and theoretical simulations. We find that the common envelope jets supernova (CEJSN) r-process scenario can account for the different observations of  r-process nucleosynthesis, including the presence of r-process-rich low-metalicity stars  in ultra-faint dwarf (UFD) galaxies. The neutron star (NS)-NS merger scenario and the collapsar scenario encounter some difficulties. Despite that, we conclude that it is very likely that more than one r-process scenario exists, with a significant contribution by the CEJSN r-process scenario. We give a prescription to include this scenario in population synthesis studies of the r-process. 
\end{abstract}


\section{INTRODUCTION}
\label{sec:intro}

Despite the significant efforts made in the field, there is no consensus on the exact site(s) where heavy r-process nucleosynthesis occurs. 
\cite{Winteleretal2012} suggested this nucleosynthesis to take place inside jets in magneto-rotational supernovae (SNe). One problem of this scenario is the high neutrino luminosity from the newly born neutron star (NS) that lowers the neutron fraction (e.g. \citealt{Pruetetal2006}) and by that substantially reduces the heavy r-process nucleosynthesis (e.g., \citealt{Papishetal2015, SokerGilkis2017}). Because of this high neutrino flux we do not study this core collapse supernova (CCSN) process here, but rather the newer proposed collapsar r-process scenario that occurs in a CCSN. 
      
A scenario that gained popularity since the kilonova event SSS17a/AT2017gfo (\citealt{Droutetal2017}; gravitational wave event GW170817, \citealt{Metzger2017}) is  binary NSs merger, where two merging NSs eject neutron-rich material, and the nucleosynthesis occurs inside the ejecta (e.g. \citealt{Rosswogetal2014}). However, there are various unsolved problems concerning this r-process scenario that brought some to suggest that more than one r-process site exists (e.g. \citealt{Coteetal2019}). 

\cite{Papishetal2015} propose and \cite{GrichenerSoker2019} further study the formation of heavy r-process elements inside the jets of common envelope jets supernovae (CEJSNe), where a cold NS accretes mass from the core of a giant star and launches neutron-rich jets. 

\cite{Siegeletal2019} study r-process nucleosynthesis in jets launched by a black hole (BH) that is formed in a collapsar event. This recently proposed scenario combines ingredients from two earlier scenarios. From the magneto-rotational SN scenario it takes the component of jets during CCSNe explosions, and from the CEJSN r-process scenario it takes the process of rapid accretion of material from the core of a massive star through a massive accretion disk. 

In the present study we critically examine the properties of the last three heavy r-process sites mentioned above (section \ref{sec:Properties}), and then compare them to recent observations and theoretical simulations (section \ref{sec:Observations}).  We summarize our results in section \ref{sec:Summary}, giving a general outline on the inclusion of CEJSN in population synthesis studies.

\section{Relevant properties of the scenarios}
\label{sec:Properties}

In Table \ref{table:properties} we list some properties of three heavy r-process scenarios that we described in section \ref{sec:intro}, the NS-NS merger, the collapsar, and the CEJSN r-process scenarios. We turn to elaborate on these properties, excluding the observational signatures that are not directly relevant for our study. 
\begin{table*}
\begin{center}
\caption{Some properties of the three r-process scenarios}
\setlength\tabcolsep{3 pt}
\begin{tabular}{|| l l l l  ||}
 \hline
  & \makecell{NS-NS merger} & \makecell{Collapsar} & \makecell{CEJSN r-process} \\ 
 \hline
\makecell{r-process yield \\  $M_{\rm rp} (M_{\rm \odot})$ $[\S \ref{subsec:Quantitaitve}] $ }& \makecell{$\approx 0.01$ \textcolor{blue}{$^{\rm [Ko12]}$} } & \makecell{$0.08-0.3$ \textcolor{blue}{$^{\rm [Si19]}$} } &  \makecell{$\approx 0.01-0.03$ \textcolor{blue}{$^{\rm [GS19]}$} }  \\
 \hline
 \makecell{Gas accretion rate \\ $(M_\odot \s^{-1})$ $[\S \ref{subsec:Quantitaitve}]$ }& \makecell{-} & \makecell{$>10^{-3}$ \textcolor{blue}{$^{\rm [Si19]}$} } & \makecell{ $\lesssim 0.1$ \textcolor{blue}{$^{\rm [GS19]}$} } \\
 \hline
 \makecell{Event number \\ $x_{\rm rp}=\frac{N_{\rm CCSN}}{N_{\rm rp}}$  $[\S \ref{subsec:Quantitaitve}]$} & \makecell{$50\leq x_{\rm rp}\leq 1350$ \textcolor{blue}{$^{\rm [Bo19]}$} \\ $x_{\rm rp}\lesssim 150$ for MW-like \\galaxies; lower rates \\for dwarf galaxies.} & \makecell{$\simeq  10^{3}-10^{4}$\\at low  \\ metalicities. \textcolor{blue}{$^{\rm [LN06]}$}} & \makecell{$\approx\rm few \times 100-$ \\ $1000$. \textcolor{blue}{$^{\rm [GS19]}$} } \\
 \hline
 \makecell{Delivery of r-process \\ elements $[\S \ref{subsec:Delivery}]$} & \makecell{Polar and equatorial \\ ejecta mix with the \\ ISM/IGM. \textcolor{blue}{$^{\rm [Bo19]}$}} & \makecell{Disk wind mixes \\ with the CCSN \\ ejecta. \textcolor{blue}{$^{\rm [Si19]}$}} & \makecell{Jets mix with \\ ISM along large \\ distances.}\\
 \hline
\makecell{Natal kick \\ $v_{\rm nk} (\km \s^{-1})$ $[\S \ref{subsec:kick}]$} & \makecell{Two natal \\ kicks lead to\\$v_{\rm nk} \sim 100$. \textcolor{blue}{$^{\rm [AZ19]}$}} & \makecell{No substantial \\ kick.} & \makecell{$v_{\rm nk} \simeq 10-20$ \\ due to one natal \\  kick. \textcolor{blue}{$^{\rm [CC13]}$}} \\
 \hline
\makecell{Delay time \\ $t_{\rm 0-rp}$ (Myr) $[\S \ref{subsec:kick}]$ }& \makecell{$t_{\rm 0-rp,NS} \simeq$ \\ $10-100$. \textcolor{blue}{$^{\rm [AZ19]}$} \\ $t_{\rm 0-rp,NS} \simeq $ \\ $40-3000$. \textcolor{blue}{$^{\rm [Co19]}$}} & \makecell{$t_{\rm 0-rp,Co} \simeq 10$.} & \makecell{$t_{\rm 0-rp,JS} \simeq $ \\ $10-30$.} \\
 \hline
\makecell{Delay distance \\ $D_{\rm nk} (\kpc)$ $[\S \ref{subsec:kick}]$}  & \makecell{$D_{\rm nk} \simeq $ \\ $1-10$. \textcolor{blue}{$^{\rm [AZ19]}$} }& \makecell{ $D_{\rm nk} \simeq 0$.} & \makecell{$D_{\rm nk} \simeq $ \\ $0.1-0.6$.} \\
 \hline
\makecell{ Observational\\ signature} & \makecell{Kilonova; Similar to \\  SSS17a/AT2017gfo \\ \textcolor{blue}{$^{\rm [e.g., Dr17]}$}.} & \makecell{SN of a very massive \\  star. Might be faint \\ or bright. Leaves\\  a BH remnant. \textcolor{blue}{$^{\rm [Si19]}$}} & \makecell{A very bright SN. \\ High-velocity and \\ highly polarized.\\ NS or BH remnant. \\ \textcolor{blue}{$^{\rm[Sc19; SG18;So19]}$}} \\ 
 \hline
\end{tabular}
\end{center}
{\textbf{Acronyms:} NS: neutron star; BH: black hole; rp: r-process; SN: supernovae; CCSN: core collapse supernovae; CEJSN: common envelope jets supernova; ISM: interstellar medium; IGM: intergalactic medium; MW: milky-way.
\\ \textbf{Definitions:} $M_{\rm rp}$: mass of newly synthesized r-process elements per event; $x_{\rm rp}$: rate of the event that produces r-process elements relative to CCSNe; $N_{\rm CCSN}$: number of CCSNe;  $N_{\rm rp}$: number of r-process events; $v_{\rm nk}$: velocity of the system due to the natal kick; $D_{\rm nk}$: distance from star formation zone to r-process production event; $t_{\rm 0-rp,NS}$/$t_{\rm 0-rp,Co}$/$t_{0-rp,JS}$: time from initial star formation to r-process enrichment by NSs merger/collapsar/CEJSN, respectively.
\\ \textbf{References:} AZ19: \cite{AndrewsZezas2019}; Bo19: \cite{Bonettietal2019}; CC13: \cite{ColeiroChaty2013}; Co19: \cite{Coteetal2019}; Dr17: \cite{Droutetal2017}; GS19: \cite{GrichenerSoker2019}; Ko12: \cite{Korobkinetal2012}; LN06: \cite{LangerNorman2006}; Sc19: \cite{Schroderetal2019}; SG18: \cite{SokerGilkis2018}; Si19: \cite{Siegeletal2019}; So19: \cite{Sokeretal2019}.  
}
\label{table:properties}
\end{table*}
  
\subsection{Yield, accretion rate, and event rate}
\label{subsec:Quantitaitve}

In the NS-NS merger scenario the destroyed NS supplies neutron-rich material to the accretion disk. In the other two scenarios the supply is of a regular gas that becomes neutron-rich in the dense accretion disk due to electron capture. The third row in Table \ref{table:properties} deals with the accretion rate of this regular gas. In the collapsar scenario large amounts of gas come from the envelope, so the accretion time is relatively long
(about hundreds of seconds). The large accreted mass yields large mass of r-process nucleosynthesis (second row, third column). 

In the CEJSN r-process scenario the accretion of mass is from the core of the primary star, and it lasts for about tens of seconds, and contains less mass and therefore less r-process nucleosynthesis (second row fourth column). In \cite{GrichenerSoker2019} we use the Bondi-Hoyle-Lyttleton mass accretion rate and find  the upper limit to be ${\rm few} \times 0.1 M_\odot \s^{-1}$. As this is an upper limit on the gas accretion rate, in this table we take it to be $\la 0.1 M_\odot \yr^{-1}$. 
This accretion rate is about two orders of magnitude larger than the accretion rate required by the collapsar scenario to form heavy r-process elements \citep{Siegeletal2019}. However, the typical yield of r-process elements in the collapsar is $\approx 10$ times bigger than in the two other scenarios, compatible with the collapsar being a longer nucleosynthesis event.  

The yield per event and the event rate of the NS-NS merger and CEJSN scenarios are similar to those that \cite{Beniaminietal2016} deduced in their model-independent study of ultra-faint dwarf galaxies (UFDs).  

In the fourth row we list the number of r-process events, $N_{\rm rp}$, as its ratio to the number of all CCSNe, $x_{\rm rp} \equiv N_{\rm rp}/ N_{\rm CCSN}$. 

We note that in all these properties, the CEJSN r-process scenario is between the two other scenarios. 

\subsection{Delivery of r-process elements}
\label{subsec:Delivery}

The r-process nucleosynthesis occurs in a fast outflow, either in jets (or disk winds) in all three scenarios, or in an equatorial outflow in the NS-NS merger scenario  \citep{Metzger2017}. Due to a natal kick (section \ref{subsec:kick}) many NS-NS merger events take place far from the place where the binary system was formed, even outside their parent galaxy, and the r-process-rich jets and equatorial outflow mix with the interstellar medium (ISM) or intergalactic medium (IGM). In a collapsar event the jets most likely mix with the ejecta of the CCSN, i.e., the previous envelope of the star, or with the cloud from which the progenitor was formed, as there is no natal kick.  
 
In the CEJSN the NS that spirals-in through the envelope accretes mass and launches jets even before it reaches the core. The jets remove the envelope, or at least clean the polar directions \citep{Sokeretal2019}. Therefore, when the NS launches jets as it accrets mass from the core, the jets expand to very large distances in the ISM.   

\subsection{The natal kick}
\label{subsec:kick}

A very important property is the distance between the r-process nucleosynthesis location and the place where the progenitors were formed. This delay distance is given by $D_{\rm nk} = v_{\rm nk} \left( t_{\rm 0-rp} - t_{\rm kick} \right)$, where $v_{\rm nk}$ is the natal kick, $t_{\rm 0-rp}$ is the delay time from star formation to r-process nucleosynthesis, and $t_{\rm kick} \simeq 10 \Myr$ is the time from star formation to the time the system suffers the natal kick. 
We list these quantities for the three scenarios is rows 6-8 of Table \ref{table:properties}. 

The NS-NS binary system suffers from the largest natal kick due to the two CCSNe, and it has the largest delay time (with a large uncertainty, e.g., \citealt{AndrewsZezas2019, Coteetal2019}). For that, the distance $D_{\rm nk}$ is very large, and might even carry it out of the galaxy (e.g. \citealt{Bonettietal2019}).  
The collapsar has no natal kick, while the CEJSN has a medium natal kick velocity as the binary system receives one natal kick while the secondary star is still massive. This is the same as the natal kick of massive X-ray binaries, $v_{\rm nk} \simeq 10-20 \km \s^{-1}$ (e.g., \citealt{ColeiroChaty2013}). 

The CEJSN r-process scenario has a typical kick velocity and a typical delay distance values in between the two other scenarios. 
   
\section{Comparison with observations}
\label{sec:Observations}

We turn to compare the three scenarios with some observations that we list in the first column of Table \ref{table:observations}. 
\begin{table*}
\begin{center}
\caption{Comparison of r-process scenarios with observations}
\setlength\tabcolsep{3 pt}
\begin{tabular}{|| c c c c ||}
 \hline
 & \makecell{NS-NS merger} & \makecell{Collapsar} & \makecell{CEJSN r-process } \\ 
  & & & \makecell{ \textcolor{blue}{$^{\rm [This~paper]}$}} \\ 
 \hline
 \makecell{ The MW \\ ${\rm [Eu/Fe]}$ evolution \\ knee $[\S \ref{subsec:knee}]$} & \textcolor{red}{\makecell{$(-)$ Expected DTD \\ ($\propto t^{-1}$ \textcolor{blue}{$^{\rm [Do12; Fo17]}$}) \\ does not reproduce \\ the ``knee''.$^\ast$ \textcolor{blue}{$^{\rm [Co19]}$} }}  & \makecell{$(+)$ Combination of \\ NS-NS mergers and \\ collapsars in the early \\ Universe can account \\ for the 'knee'. \textcolor{blue}{$^{\rm [Si19]}$}} & \makecell{$(+)$ R-process during \\ CEE with short delay \\ after star formation.} \\
 \hline
 \makecell{Short-lived \\ radioactive \\ isotope in the \\ early Solar \\ System $[\S \ref{subsec:SLR}]$} & \makecell{$(+)$ Compatible with \\ observations. \textcolor{blue}{$^{\rm [BM19]}$} } &   \textcolor{red}{\makecell{$(-)$ Events too rare \\ at Solar System\\  formation, \\ $t_{\rm MW}= 9 \Gyr$. \textcolor{blue}{$^{\rm [BM19]}$}}}  & \makecell{$(+)$ At $t_{\rm MW}= 9 \Gyr$ \\ the event rate is\\ similar to that \\ of NS-NS merger.} 
 \\
 \hline
 \makecell{No correlation  \\ of r-process \\ and iron in \\ the  young \\  MW $[\S \ref{subsec:Iron}]$} & \makecell{$(+)$ The delay distances \\ $D_{\rm nk} \simeq 1-10 \kpc$ \textcolor{blue}{$^{\rm [AZ19]}$} \\  are sufficiently large \\ to dilute the \\ r-process elements in \\ a large volume. \textcolor{blue}{$^{\rm [Sh15]}$} } & \textcolor{red}{\makecell{ $(-)$ $D_{\rm nk} \simeq 0$ implies \\ mixing of the r- \\ process elements with \\ iron that the collapsar \\ forms, correlating iron \\ with r-process. \textcolor{blue}{$^{\rm [MRR19]}$} }}  & \makecell{$(+)$ A moderate kick \\ $+$ jets' propagation \\ dilute the r-process \\ in a large volume.} \\
 \hline
 \makecell{Presence of the \\ r-process \\ elements inside \\ UFD galaxies of \\ radius $R_{\rm DG}$ $[\S \ref{subsec:KickDwarf}]$} & \textcolor{red}{\makecell{ $(-)$ $D_{\rm nk} > R_{\rm DG}$, \\ implies the \\ distribution of \\ r-process elements \\ mainly outside UFD\\ galaxies.$^\ast$ \textcolor{blue}{$^{\rm [BL16; Bo19]}$} }} & \makecell{$(+)$ R-process in \\ star forming \\ regions inside \\ the UFD galaxy. \\ } & \makecell{ $(+)$ $D_{\rm nk} \lesssim R_{\rm DG}$ \\ leaves most \\ r-process \\ elements inside  \\ the UFD galaxy. \\ } \\
 \hline
 \makecell{R-process within \\ a short timescale \\ $t_{\rm DG} < 1 \Gyr$ \\ of star formation \\ in UFD galaxies \\ $[\S \ref{subsec:TimescaleDwarf}]$} & \textcolor{red}{\makecell{$(-)$ $t_{\rm 0-rp,Ns} > t_{\rm DG}$, \\ implies too low \\ r-process enrichment \\ of the oldest stars.$^\ast$ \textcolor{blue}{$^{\rm [Bo19]}$} }}  & \makecell{$(+)$ $t_{\rm 0-rp,Co} < t_{\rm DG}$ \\ allows r-process \\ enrichment of \\ oldest stars. } & \makecell{$(+)$ $t_{\rm 0-rp,JS} < t_{\rm DG}$ \\ allows r-process \\ enrichment of \\ oldest stars.  }\\ 
 \hline
\end{tabular}
\end{center}
{\textbf{Acronyms:} NS: neutron star; SN: supernova; CCSN: core collapse supernova; CEJSN: common envelope jets supernova; MW: milky-way; DTD: delay time distribution; CEE: common envelope evolution; ISM: interstellar medium; UFD: ultra-faint dwarf.
\\ \textbf{Definitions:} $D_{\rm nk}$: distance from star formation zone to r-process distribution location; $R_{\rm DG}$: typical radius of a UFD galaxy; $t_{\rm DG}$: formation time of the oldest stars in UFD galaxies.
\\ \textbf{References:} AZ19: \cite{AndrewsZezas2019}; BL16: \cite{BramanteLinden2016}; BM19: \cite{BartosMarka2019}; Bo19: \cite{Bonettietal2019}; Co19: \cite{Coteetal2019}; Do12: \cite{Dominiketal2012}; Fo17: \cite{Fongetal2017}; Ji16: \cite{Jietal2016}; MRR19: \cite{MaciasRamirez-Ruiz2019}; Ro16: \cite{Roedereretal2016}; Sh15: \cite{Shenetal2015}; Si19: \cite{Siegeletal2019};  \\
$^\ast$ See section \ref{subsec:BH} for possible solution. 
}
\label{table:observations}
\end{table*}

\subsection{The ${\rm [Eu/Fe]}$ evolution ``knee''}
\label{subsec:knee}

The ${\rm [Eu/Fe]}$ evolution ``knee'' in the plane of ${\rm [Eu/Fe]}$ versus ${\rm [Fe/H]}$ refers to the change of behavior around  ${\rm [Fe/H]} \simeq -1$. For low metalicity stars with ${\rm [Fe/H]} \la -1$ there is a large scatter in the value of ${\rm [Eu/Fe]}$ around about the same average value, while stars of ${\rm [Fe/H] \ga -1}$ show a trend of decreasing value of  $\rm [Eu/Fe]$ with increasing value of $\rm [Fe/H]$. This suggests a decreasing production of r-process elements when the metallicity in the Milky Way was ${\rm [Fe/H]} \simeq -1$, corresponding to a Galaxy age of $\lesssim 1 \Gyr$. 
\citealt{Coteetal2019} argue that the delay time distribution (DTD) of merging NSs, that is similar to that of gamma ray bursts (GRBs; e.g. \citealt{Simonettietal2019}), is such that the merger rate decreases too slowly to account for the ${\rm [Eu/Fe]}$ evolution ``knee''. 
They discuss a solution where an additional r-process site existed in the early Universe. This site faded faster, before the the onset of the first type Ia SNe that produced large amounts of iron. 

The delay time from star formation to NS-NS merger, $t_{\rm 0-rp, NS}$, includes the time to form the binary NS system, and then the much longer gravitational radiation time to merger.  
\cite{Coteetal2019} show that a steeper DTD than that of the NS-NS merger can account for the ${\rm [Eu/Fe]}$ evolution ``knee''. For example, a delay time where all r-process events occur within $<10^8 \yr$ from star formation is suitable for explaining the ``knee''. Both the collapsar \citep{Siegeletal2019} and the CEJSN r-process scenario have a much shorter delay time from star formation to r-process nucleosynthesis than the NS-NS merger scenario (Table \ref{table:properties}), as they include only the stellar evolution time of massive stars that is $\ll 10^8 \yr$. 
Moreover, \cite{Safarzadehetal2019b} found that NS-NS mergers cannot account for r-process enrichment at early times given the delay times that are listed in table \ref{table:properties}. 
We therefore mark in Table \ref{table:observations} that both these scenarios can explain the ``knee'', while the NS-NS merger scenario cannot (in red). \cite{BeniaminiPiran2019} note, however, that a steeper decline of NS-NS merger rate than the commonly accepted $t^{-1}$ might explain the ${\rm [Eu/Fe]}$ evolution ``knee''.

\subsection{Short-lived radioactive isotopes in the early Solar System}
\label{subsec:SLR}

\cite{BartosMarka2019} argue that because the event rate of collapsars decays rapidly with the age of the Galaxy, at the time of Solar System formation when the Milky Way (MW) age was $t_{\rm Mw} \simeq 9 \Gyr$, the event rate was already too low to account for short-lived radioactive isotopes that are synthesized in the r-process. We therefore mark the collapsar scenario with red in the third row of Table \ref{table:observations}. \cite{BartosMarka2019} further argue that the NS-NS merger scenario does not suffer from this problem. 
 
The time distribution of the CEJSN r-process scenario resembles the time distribution of the formation of NS binary systems. However, the NS-NS merger scenario has additional delay from NS binary formation to merger due to gravitational wave radiation. This time scale is $t_{\rm GW} \approx 0.01-1 \Gyr$. As  $t_{\rm GW} \ll t_{\rm MW}$, at the time of Solar System formation the event rate of the CEJSN r-process was not much different than that of the NS-NS merger. We therefore suggest here that the CEJSN r-process scenario can account for the short-lived radioactive isotopes at Solar System formation. 

\subsection{No correlations with iron production}
\label{subsec:Iron}

\cite{MaciasRamirez-Ruiz2019} find that the collapsar scenario overproduces r-process elements relative to iron in metal poor stars in the MW. This is because the collapsar does not have a natal kick and so the r-process production occurs in the same region where the iron is synthesized due to the CCSN explosion.
They argue that the values of $\rm [Eu/Fe]$ in metal poor stars suggest that the r-process site is spatially uncorrelated with iron production. In other words, the $\rm [Eu/Fe]$ chemical evolution pattern was produced by diluting the r-process elements by the time the r-process-rich gas mixes with iron-rich gas that was produced in other sites, i.e., not that of the r-process.  

The NS-NS merger scenario fulfills this condition as the high natal kick velocity and long delay time bring the NS binary system far from its birthplace before it merges. 
However, \cite{MaciasRamirez-Ruiz2019} note that the NS-NS merger scenario still has difficulty explaining the ${\rm [Eu/Fe]}$ evolution ``knee'' (section \ref{subsec:knee}). 

The natal kick of the binary system in the  CEJSN r-process scenario is between that of the two other scenarios (Table \ref{table:properties}). This is sufficient to bring most of the systems outside their star formation zone when the r-process nucleosynthesis takes place, and ensure no spatial correlation between the r-process and iron production sites, as required by \cite{MaciasRamirez-Ruiz2019}.

\cite{Beniaminietal2018} suggest that if a significant portion of extremely metal poor stars in the Galaxy comes from UFD-like galaxies that have long ago been dissolved to form the galactic halo stellar population, then the mean value and scatter of ${\rm [Eu/Fe]}$ as a function of ${\rm [Fe/H]}$ in these stars can be explained in the frame of the NS-NS merger scenario. We note that the same explanation would apply for the CEJSN r-process scenario. 

\subsection{Presence of r-process elements inside ultra faint dwarf galaxies}
\label{subsec:KickDwarf}
 
The metal poor stars of the UFD galaxy Reticulum II contain r-process elements (\citealt{Jietal2016, Roedereretal2016}). The natal kick velocity of the NS-NS binary system (section \ref{sec:Properties}) is typically an order of magnitude and more above the stellar dispersion velocity of Reticulum II, $\sigma \simeq 3.3 \km \s^{-2}$ (e.g., \citealt{Simonetal2015}), and due to the long delay time to merger the r-process will occur mainly outside such a galaxy (see early discussion by \citealt{SafarzadehScannapieco2017}).
If NS-NS mergers were the only source of heavy r-process elements, then UFD galaxies should not have a signature of heavy elements, and therefore another nucleosynthesis site of heavy r-process elements is required \citep{Bonettietal2019}. 
An alternative explanation in the frame of the NS-NS merger scenario is a slow natal kick 
\citep{BeniaminiPiran2016} and the delivery of r-process elements to large distances by the ejecta of the merger \citep{Beniaminietal2018}. Here we take the recent view of \cite{Bonettietal2019} that there is indeed a problem with the NS-NS merger scenario due to higher kick vlocities.  

The two other scenarios we consider do not have this problem. 
The collapsar does not experience a natal kick and the r-process takes place in the parent star forming zone (that raises other problems; see section \ref{subsec:Iron}), and therefore inside the hosting UFD galaxy. 
The CEJSN r-process scenario has a moderate kick velocity that brings most of the systems out from their original star forming clouds (section \ref{subsec:Iron}), but the moderate natal kick velocity and the short delay time make the typical delay distance $D_{\rm nk}$ smaller or about equal to the extent of the stellar distribution in UFD galaxies.

\subsection{R-process in the oldest stars of UFD galaxies}
\label{subsec:TimescaleDwarf}

\cite{Bonettietal2019} raise another problem of the NS-NS merger scenario in regards to r-process in UFD galaxies. The old stellar population in these galaxies is thought to be the result of a fast star formation episode, that ends within the first Gyr of the galactic evolution. \cite{Bonettietal2019} find that the long lasting DTD of the NS-NS merger scenario (e.g. \citealt{Coteetal2019}) is such that many events take place too late to account for the r-process abundance in old stellar populations in UFD galaxies. \cite{Safarzadehetal2019a}, on the other hand, argue that fast merging double NSs can account for r-process elements in UFD galaxies in terms of merger locations and timescales. 
The delay times of the collapsar and the CEJSN r-process scenarios (Table \ref{table:properties}) are much shorter than for the NS-NS merger, and can unequivocally account for r-process elements in these very old stellar populations. 

\subsection{Replacing a NS with a BH}
\label{subsec:BH}
 
Replacing one of the NSs in some events of the NS-NS merger scenario with a BH might ease the problems we mention in Table \ref{table:observations}. \cite{Wehmeyeretal2019} consider BH-NS merger as an r-process production site in addition to NS-NS merger. They argue that the BH-NS merger rate is higher in the early Universe and decays more rapidly than NS-NS merger rate. This behavior better fits the  ${\rm [Eu/Fe]}$ evolution ``knee'' in the MW (section \ref{subsec:knee}) and the r-process abundance in the old stellar populations in UFD galaxies (section \ref{subsec:TimescaleDwarf}). We note here that the natal kick velocity of the more massive BH-NS binary system can ease the problem of the presence of r-process elements in UFD galaxies (section \ref{subsec:KickDwarf}). 
However there are major caveats in this BH-NS scenario. The BH cannot be more massive than about $10-14 M_\odot$, and it should have large enough angular momentum to ensure ejection of mass by the accretion disk. 

We propose here the possibility of replacing the NS companion in some of the CEJSN r-process scenario events with a BH. We note that for solar metaliciteis, a merger of a BH with the core might be almost as common as the merger of a NS with the core \citep{Schroderetal2019}. This result would not vary much for lower metalicites.

Like with NS-BH binary systems that are formed in larger relative numbers in the early Universe, here also the event rate, and as a result the r-process yield, would be larger in earlier times. 

In addition, the neutrino emission from the mass-accreting BH is lower than in the case of a mass-accreting NS, therefore reducing the conversion of neutrons to protons.  \cite{Bonettietal2019}, for example, claim that when a central long-lived massive NS is formed in a NS-NS merger, the large flux of neutrinos that convert neutrons to protons prevent heavy r-process nucleosynthesis. 
However, if the central NS collapses immediately to a BH (in a timescale that is shorter than the disc viscous timescale), then heavy elements will be synthesized, as the material that is accreted onto the BH is not cooled by neutrino emission.
The same reasoning applies also to a NS or a BH companion in the CEJSN r-process scenario. 
We note that since the core of the giant star from which the BH accretes mass in this modified CEJSN r-process scenario is very large, the constraints on the BH properties that exist in the case of the NS-BH merger scenario \citep{Wehmeyeretal2019} do not exist here.

The study of the CEJSN r-process scenario with a BH companion is the subject of a forthcoming paper. 

\section{How to include the CEJSN r-process in population synthesis}
\label{sec:Summary}
 
We studied the basic properties of three different possible heavy r-process scenarios (sites; section \ref{sec:Properties}) and compared them to observations (section \ref{sec:Observations}). We found that the CEJSN r-process scenario has some advantages over the NS-NS merger scenario and over the collapsar scenario (Table \ref{table:observations}). Although we list no problems with the CEJSN r-process scenario, it is very likely that more than one scenario exists. Table \ref{table:observations} brings us to conclude that the CEJSN r-process scenario has a significant contribution to the r-process nucleosynthesis, especially in the early Universe where NS-NS mergers could not account for the formation of the heavy elements. 
 
Including the CEJSN r-process scenario would be valuable for population synthesis models.
The most important property of this scenario to consider is that the evolution of the CEJSN r-process scenario is almost identical to the evolution to form NS-NS binary system, but there is no time delay due to gravitational wave emission until merger.
Therefore, the inclusion of the CEJSN r-process scenario in r-process population synthesis studies should include the following modifications with respect to the NS-NS merger scenario. 
\begin{enumerate} 
\item The rate of the CEJSN r-process scenario is about $0.3-1$ times that of the NS-NS merger event rate. 
\item The r-process yield per event, $M_{\rm rp}$, is $1-3$ times larger than that in the NS-NS merger event. 
\item The DTD is like the time distribution for the formation of NS-NS binary. The typical time from star formation to the CEJSN r-process event is $t_{\rm 0-rp,JS} \simeq 10-30 \Myr$. Namely, the difference from the DTD of the NS-NS merger scenario is that there is no delay due to gravitational waves. 
\item The distance of each r-process event from its parent star formation location should be calculated with a smaller kick velocity of only $v_{\rm nk} \simeq 10-20 \km \s^{-1}$.   
\end{enumerate}

\end{document}